\documentclass[%
 aip,
 jmp,%
 amsmath,amssymb,
 reprint,%
]{revtex4-1}

\usepackage{graphicx}
\usepackage{listings}
\usepackage{dcolumn}
\usepackage{color}

\definecolor{cdarkblue}{rgb}{0.1,0.1,0.5}
\definecolor{cdarkred}{rgb}{0.5,0.1,0.1}
\definecolor{red}{rgb}{1.0,0,0}
\definecolor{dgreen}{rgb}{0.3,0.8,0.3}

\newcommand{\ben}{\begin{enumerate}}
\newcommand{\een}{\end{enumerate}}
\newcommand{\bit}{\begin{itemize}}
\newcommand{\eit}{\end{itemize}}
\newcommand{\be}{\begin{equation}}
\newcommand{\ee}{\end{equation}}
\newcommand{\bdm}{\begin{displaymath}}
\newcommand{\edm}{\end{displaymath}}
\newcommand{\bea}{\begin{eqnarray}}
\newcommand{\eea}{\end{eqnarray}}
\newcommand{\f}[1]{\fbox}

\newcommand{\vel}{v}

\draft 

\begin{document}


\title{ Random Numbers from a Delay Equation}

\author{Julian Self}
\email[Author to whom correspondence should be addressed:]{julian.self@mail.mcgill.ca}
\affiliation{Departments of Physics, and Physiology,
and Centre for Applied Mathematics in Bioscience and Medicine (CAMBAM),
McGill University, 3655 Promenade Sir William Osler, Montr$\acute{e}$al, QC, Canada H3G 1Y6}

\author{Michael C. Mackey}
\email[]{michael.mackey@mcgill.ca}
\affiliation{Departments of Physiology, Physics, and Mathematics,
and Centre for Applied Mathematics in Bioscience and Medicine (CAMBAM),
McGill University, 3655 Promenade Sir William Osler, Montr$\acute{e}$al, QC, Canada H3G 1Y6}

\date{\today}

\begin{abstract}
Delay differential equations (DDE) can have ``chaotic" solutions that can be used to mimic Brownian motion. Since a Brownian motion is random  in its velocity, it is reasonable to think that a random number generator (RNG) might be constructed from such a model. In this preliminary study, we consider one specific example of this and show that it satisfies criteria commonly employed in the testing of random number generators (from TestU01's very stringent ``Big Crush" battery of tests). A technique termed digit discarding, commonly used in both this generator and physical RNG's using laser feedback systems, is discussed  with regard to the maximal Lyapunov exponent. Also, we benchmark the generator to a contemporary common method: the multiple recursive generator, MRG32k3a. 
Although our method is  about 7 times slower than MRG32k3a, there is in principle no apparent limit on the number of possible values that can be generated from the scheme we present here.
\end{abstract}

\maketitle

\begin{quotation}
Deterministic differential delay equations are well known to sometimes have chaotic solutions that are unpredictable in spite of the fact that they approach either ensemble or trajectory limiting densities that are independent of initial conditions (functions). We show that this characteristic may be used effectively for producing a random number generator.

\end{quotation}

\section{Introduction}\label{sec:intro}
\label{sec:introduction}
From Monte Carlo simulators to student selection in American charter schools to financial transactions, random number generators (RNG) are widely employed. It is difficult to articulate what constitutes numbers that are truly random, but often, if generators pass a defined battery of tests, they are said to be random.

In this paper, we show how a first order differential equation with a delayed argument (differential delay equation, DDE) that has been recently studied can be used as an effective random number generator. In Section \ref{sec:history}, a very brief history of popular RNG's is given. In Section \ref{sec:chaotic}, a previously studied DDE producing a Brownian motion is introduced. Section \ref{sec:DDE-RNG} introduces a straight-forward scheme for generating numbers from a DDE. Section \ref{sec:LSF} discusses how to increase generation speed by borrowing a technique from comparable and experimentally realized feedback laser systems. This technique, termed digit discarding, and its potential relationship to the Lyapunov exponent are discussed. Section \ref{sec:benchmarking} contains a comparison of the DDE as a RNG and a standard generator method, the Multiple Recursive Generator.

This is, as far as the authors know, the first random number generator to employ a differential delay equation, while producing high quality random numbers. The quality of the numbers is matched by but a few documented generators, with a period no shorter than any other. It is important to remember that when employing random numbers, one does not a priori know the result of, say, a given simulation, so it is impossible to say in which way it would be acceptable for a generator to be systematically flawed. Furthermore, while periods exceeding currently used generators' cannot be readily shown to be needed for a stream, longer periods are typically seen as being tied to higher quality generated numbers. \cite{ECUYER}

\section{A Very Short History of RNG's}\label{sec:history}


Although there also exist nondeterministic, physically implemented, RNG's,\cite{HOCS} this section focuses specifically on deterministic software RNG's. Two of the most currently used general purpose RNG's have a rich history that can be traced back to the earlier RNG's counterparts from which they were derived. The Mersenne twister is heavily inspired from the linear feedback shift register (LFSR), while the combined multiple recursive algorithm (CMRG), has its origins in the linear congruential generator. A very brief overview is presented here, and both Knuth and L'Ecuyer have given complete and detailed histories of these generators. \cite{KNUTH,HOCS}

\subsection{Linear congruential generators}
The linear congruential generator (LCG) was introduced in 1949 by D.H. Lehmer, \cite{KNUTH} in which, for integers $X_n$, the following sequence can be expressed:
\begin{equation}
\label{eq:LCG}
X_n=(X_{n-1}a+c)~mod~m
\end{equation}

The modulus is denoted $m$, the multiplier $a$, the increment $c$ and the starting value $X_0$.\cite{KNUTH} The random number output $U_n$ can be obtained by dividing $X_n$ by $m$. Much work has been done on studying what values the multiplier, increment and modulus must have for better distributed output sequences and longer periods. For example, the period length can only be of length $m$ if the increment is relatively prime to the modulus. These generators are still used today, for example they are the default RNG in Java. The output sequences do possess serious flaws in their structure, and so are not suggested.

LCG's were later generalized to multiple recursive generators (MRG), where $X_n$ is a function of not only $X_{n-1}$, but of linear combinations of $(X_{n-1},...,X_{n-k})$. So-called lagged Fibonacci generators are of this type.

The MRG algorithm was further improved by employing different MRG's in parallel to form the input of a new modular recurrence relation for the aptly called combined multiple recursive generator (CMRG). This latter generator provides sequences much better distributed than its antecedent, the MRG. The details of the CMRG can be found in L'Ecuyer, and one widely used implementation is the MRG3k32a. \cite{ECUYER}
\subsection{Linear feedback shift registers}

In 1965 Tausworthe introduced a binary representation RNG utilizing a recurrence relation modulo 2. \cite{TAUSWORTHE} It can be expressed by the following relation:\cite{HOCS}

\begin{equation}
\label{eq:dc}
\begin{array}{cc}
X_i = (c_1 X_{i-1} + ... + c_k X_{i-k})~mod~2 \\
U_i =\sum_{l=1}^{w} X_{is+l-1}2^{-l}
\end{array}
\end{equation}

In this equation, ${\bf c}$ and $s$ are characteristic for a given generator, $w$ is the size of the output vector and $U_i$ is a final output of this generator which is called the linear feedback shift register (LFSR).

\begin{equation}
{\bf X}_{l+n}={\bf X}_{l+m}~xor~{\bf X}_l~{\bf A}~(l=0,1,...)
\end{equation}

For ${\bf A}$ as the identity matrix, the above equation describes the generalized feedback shift register (GFSR).\cite{LEWIS,MATSUMOTO} In this case, ${\bf X}_l$ is a word of size $w$ with components $0$ or $1$ while $xor$ refers to the bitwise exclusive-or operation. The word, considered as real number between 0 and 1 in binary representation, is the pseudorandom output of the GFSR.\cite{MATSUMOTO} The GFSR was further generalized, or twisted, by picking a non-identity matrix ${\bf A}$. This finally gives rise to the twisted generalized feedback shift register (TGFSR). A variation of the TGFSR is the Mersenne twister, one implementation being MT19937,\cite{MT} which is perhaps the most widely used generator today. For example, it is the default generator in the applied mathematics software package Matlab.

\subsection{Other generators}

There are a wide variety of other RNG algorithms that have been suggested. For example, the LCG can be generalized to a non-linear recurrence relation.\cite{KNUTH}\cite{NLM} Some cryptographic cyphers may also be used as RNG's, and in some cases have been thoroughly tested.\cite{TESTU01} However, the tests commonly applied to pass cryptographic standards, e.g. the NIST tests, are weak \cite{TESTU01} and so each algorithm would have to be tested and considered separately before it could be recommended as a robust ``general purpose" RNG.

\section{Chaotic Solutions to a Delay Differential Equation}\label{sec:chaotic}

\begin{figure*}[htbp]
\centering
\includegraphics[width=16cm]{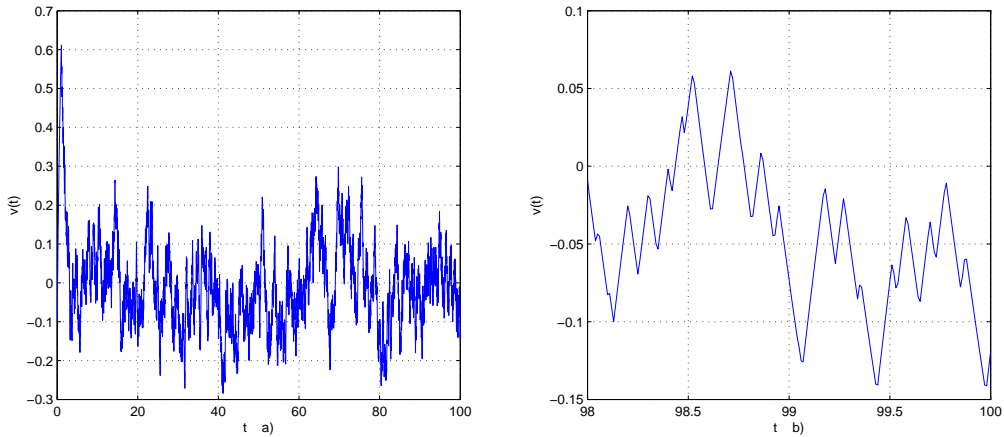}
\caption{  a) A sample solution of equation $\ref{eq:ex-2}$ with $\beta=10, \gamma = 1, f_0 = 1$, and an initial function $\phi(t) \equiv -0.1, t \in [-1,0]$. b) the solution segment for $98 \leq t \leq 100$.}
\label{fig:ex}
\end{figure*}

 {Several investigators \citep{Beck91,Chew:02, Mackey06} have shown that a Brownian-like motion can arise when a   particle is subjected to impulsive kicks $f(t)$ derived from a discrete time dynamical system, and whose dynamics are modeled by the following equations where $x$ is position, $v$ is velocity, $m$ is mass and $\gamma$ the friction coefficient: }

\begin{equation*}
\label{eq:dc}
\left\{
\begin{array}{rcl}
 \dfrac{d x}{d t} &=& \vel\\
m \dfrac{d \vel}{d t} &=& - \gamma \vel + f(t).
\end{array}\right.
\end{equation*}

Lei and Mackey\cite{DBM} sought an alternative continuous time description  of the ``random force'' $f(t)$, which was assumed to depend on the state (velocity) of a particle, but with a lag time $\tau$, i.e.,
\begin{equation*}
f(t) = F(v(t-\tau)),
\end{equation*}
where $F$ has the appropriate properties to generate chaotic solutions.  They  considered the  following differential delay equation
\begin{equation}
\label{eq:dde-0}
\begin{array}{l}
\left\{\begin{array}{rcl}
\dfrac{d x}{dt}&=& \vel\\
m\dfrac{d \vel}{dt} &=& - \gamma\vel + F(\vel(t-\tau)),
\end{array}\right.\\
\quad v(t)=\phi(t), \,\,-\tau \leq t \leq 0,
\end{array}
\end{equation}
where $\phi(t)$ denotes the initial (or history) function which must always be specified for a differential delay equation.

First some observations about the second equation in $\ref{eq:dde-0}$ which determines the dynamics of the velocity. A simple form of the ``random'' force is binary and fluctuates between $\pm f_0$, for instance given by
\begin{equation}
\label{eq:dde-heav}
F(v) = 2 f_0\left[H(\sin (2\pi \beta v)) - \dfrac{1}{2}\right],
\end{equation}
where $H$ is the Heavyside step function
\begin{equation*}
H(\vel) = \left\{\begin{array}{ll}
0\quad & \mathrm{for}\ \vel<0\\
1 & \mathrm{for}\ \vel\geq 0.
\end{array}\right.
\end{equation*}
Then we have the following equation
\begin{equation}
\label{eq:ex-2}
\dfrac{d\vel}{dt} = -\gamma\vel + 2  \left[H(\sin(2 \pi \beta \vel(t-1)) - \frac{1}{2})\right].
\end{equation}
(Here and later we always assume the mass $m=1$ and $f_0 = 1$ which can be achieved through the appropriate scaling.)
{The delay differential equation $\ref{eq:ex-2}$ with a binary ``random force'' can be solved iteratively by the method of steps. \footnote{A solution of equation $\ref{eq:ex-2}$ is associated with a time sequence $t_0 < t_1 < \cdots< t_n< \cdots$,
which is defined such that $\sin(2\pi\beta \vel(t)) \geq 0$ when $t\in [t_{2k}, t_{2k+1})$, and
$\sin(2\pi\beta \vel(t)) < 0 $ when $t\in [t_{2k-1}, t_{2k})$. Furthermore, if the sequence $(t_0, \cdots, t_n)$ is known, then the solution $\vel(t)$ when $t\in (t_n, t_n + 1)$ can be obtained explicitly, and therefore, $t_{n+1}$, which is defined as $\sin(2\beta \vel(t_{n+1})) = 0$,  is determined by $(t_0, \cdots, t_n)$. Once we obtain the entire sequence $\{t_n\}$, the solution of equation $\ref{eq:ex-2}$ consists of exponentially increasing or decreasing segments on each interval $[t_n, t_{n+1}]$. Nevertheless, the nature and properties of the  map $t_{n+1} = F_n(t_0, t_1,\cdots, t_{n})$ is still not characterized and has defied analysis to date.
} Despite its simplicity, it can display behaviours similar to a random process. An example solution of equation $\ref{eq:ex-2}$ is shown in Figure $\ref{fig:ex}$}. The ``random force'' in equation $\ref{eq:ex-2}$ is discontinuous and gives a continuous zigzag velocity curve (c.f. Figure \ref{fig:ex} b)

\subsection{ Deterministic Brownian motion}
Lei and Mackey\cite{DBM} focused on an analogous different delay equation
\begin{equation}
\label{eq:1}
\begin{array}{l}
\dfrac{d \vel}{dt} = - \gamma\vel + \sin(2\pi\beta \vel(t-1)), \\
v(t)=\phi(t), \,\,-1 \leq t \leq 0.
\end{array}
\end{equation}
In equation $\ref{eq:1}$, $\beta$ measures the ``frequency'' of the dependence of the  nonlinear function on $v(t-1)$, and this turned out to be an essential parameter in their study.  Thus they studied the dynamical properties of the solutions of equation $\ref{eq:1}$, both analytically and numerically, but really  focused on the probabilistic properties of the chaotic solutions of
\begin{equation}
\label{eq:bm}
\begin{array}{l}
\left\{\begin{array}{rcl}
\dfrac{d x}{dt}&=& \vel\\
\dfrac{d \vel}{dt} &=& - \gamma\vel + \sin(2\pi\beta \vel(t-1)),
\end{array}\right.\\
\quad v(t)=\phi(t), \,\,-1 \leq t \leq 0,
\end{array}
\end{equation}
and characterized the statistical solution properties. {Their main result was to show that equation $\ref{eq:bm}$ can reproduce experimentally observed Brownian motion data over a wide range of time scales, in spite of the fact that the evolution equation is deterministic. Therefore, the chaotic solutions of $\ref{eq:bm}$ are a deterministic Brownian motion.}

Throughout Lei and Mackey\cite{DBM}, the probabilistic properties of solutions of equations $\ref{eq:ex-2}$ and $\ref{eq:1}$ were studied numerically. In their  simulations, for a given set of parameters, they solved one of the equations  with a randomly selected constant initial function
\begin{equation*}
\label{eq:ini}
\vel(t) = \vel_0 \in (-1, 1),\quad (-1 \leq t\leq 0),
\end{equation*}
where $\vel_0$ is drawn from a uniformly distributed density.
The solution $\vel(t)$ was obtained using Euler's method (with a time step $\Delta t = 0.001$) up to $t=10^5$, and was sampled every $10^3$ steps to generate a time series $\{\vel_n\}$, where $\vel_n = \vel(n\times 10^3 \Delta t)$, and $n = 1,2,\cdots$. The resulting time series of values $\{\vel_n\}$ was used to characterize the statistical properties of the solution.

In particular, Lei and Mackey\cite{DBM} focused  on the mean value
$\mu$, the upper bound $K$, the standard deviation $\sigma$, and the excess kurtosis $\gamma_2$ of the time series, defined by
\begin{equation*}
\label{eq:pp}
\begin{array}{c}
\displaystyle
\mu = \dfrac{1}{N}\sum_{n=1}^N \vel_n,\ K = \max_n |\vel_n|,\ \sigma^2 = \dfrac{1}{N}\sum_{n=1}^N (\vel_n - \mu)^2,\\
\displaystyle
\gamma_2 = \dfrac{\mu_4}{\sigma^4} - 3,\ \mathrm{where}\ \mu_4 = \dfrac{1}{N}\sum_{n=1}^N (\vel_n - \mu)^4.
\end{array}
\end{equation*}
The excess kurtosis $\gamma_2$ measures the sharpness of the density of the sequence, and a value of $\gamma_2=0$ is characteristic of a normal Gaussian distribution.

Lei and Mackey\cite{DBM} found that for equation 
$\ref{eq:1}$ their numerical results could be approximately fit by the functions
\begin{eqnarray}
\label{eq:K0}
K(\beta,\gamma) &=& \dfrac{1}{\sqrt{\gamma} (0.68\sqrt{\beta} + 0.60 \sqrt{\gamma})}\\
\label{eq:sigma}
\sigma(\beta, \gamma) &=& \dfrac{0.32}{\sqrt{\beta \gamma}}\\
\label{eq:kur}
\gamma_2(\beta,\gamma) &=& - \dfrac{\gamma}{\beta}.
\end{eqnarray}
\begin{figure*}[htbp]
\centering
\includegraphics[width=16cm]{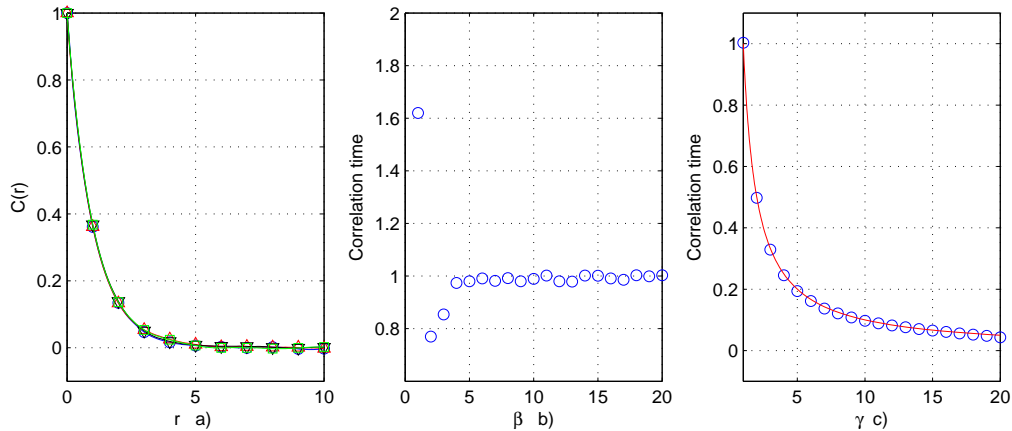}
\caption{ (a) Correlation function $C(r)$ computed for the solutions of equation $\ref{eq:1}$.  Here, $\gamma =   1$, and $\beta = 5$ (blue circles), $10$ (red up triangles), $15$ (black down triangles), $20$ (green squares), respectively. (b) Correlation time as a function of $\beta$ (with $\gamma = 1$).  (c) Correlation time as a function of $\gamma$ (with $\beta = 20$), solid curve is the fit  with $t_0 =
1/\gamma$.}
\label{fig:corr}
\end{figure*}
Additionally, they examined the behaviour of the normalized correlation function of a solution defined as
\begin{equation*}
C(r) = \lim_{T\to\infty}\dfrac{\int_0^T \vel(t)\vel(t+r) d t}{\int_0^T \vel(t)^2 d t}.
\end{equation*}
Figure \ref{fig:corr}a shows $C(r)$ for different values of $\beta$ (with $\gamma = 1$) for equation \eqref{eq:1}. From Figure \ref{fig:corr}, the correlation function can be approximated as an exponential function of the form
\begin{equation}
\label{eq:corr}
C(r) \simeq e^{-r/t_0},
\end{equation}
where $t_0$ is the correlation time. Figures \ref{fig:corr}b-c show that the correlation time is largely independent of $\beta$, and that it is approximately given by $1/\gamma$.  Identical results were found for equation $\ref{eq:ex-2}$ but they did not show these results.

From their numerical results it was clear that the excess kurtosis $\gamma_2$ of the irregular solutions of $\ref{eq:ex-2}$ and   $\ref{eq:1}$ varied with $\beta$ and $\gamma$ according to $\gamma_2 \simeq -\gamma/\beta$. Thus, the  corresponding
distributions approached Gaussian-like distributions when $\beta$ is large (and $\gamma$ is fixed), but one with a truncated tail so that it is supported on a set of finite measure. They called such truncated Gaussian distributions \textit{quasi-Gaussian distributions}.

Let $\mu$ and $\sigma$ be the mean and standard deviation of a quasi-Gaussian noise, and assume that the noise signal is supported on an interval $[\mu - K, \mu + K]$.  Then the density function is given by
\begin{equation}
\label{eq:den1}
p(\vel; \mu, \sigma, K) = \left\{
\begin{array}{ll}
\mathcal{C}_0 e^{-\frac{(\vel-\mu)^2}{2\sigma^2}}
& \qquad \mathrm{if} \quad |\vel-\mu|\leq K\\
0 & \qquad   \mathrm{ other\ wise,}
\end{array}\right.
\end{equation}
where
\begin{equation*}
\mathcal{C}_0= \dfrac{1}{(\Phi(K/\sigma) - \Phi(-K/\sigma))}
\end{equation*}
and
\begin{equation*}
\Phi(z) = \int_{-\infty}^z e^{-s^2/2} d s = \sqrt{\dfrac{\pi}{2}}\left[1 +\mathrm{erf}\left(\dfrac{z}{\sqrt{2}}\right)\right].
\end{equation*}

\section{DDE-RNG}\label{sec:DDE-RNG}
\subsection{Mapping to random numbers}\label{ssec:map}

Knowing the density, or distribution, of solutions from equation $\ref{eq:1}$, it is possible to generate random numbers. One way to to this is the following. First, $\gamma$ can be scaled to $1$, while the parameter $\beta$ should be chosen to be larger than $20$ to assure a non-periodic time series solution.\cite{DBM} The history function $\phi$ can be taken as any constant in the interval $(-1,1)$,  as it was in Lei and Mackey.\cite{DBM} Finally the Euler Method can be used with a time step of $\Delta t =  0.001$ and the time series can be sampled with an appropriate interval for a sufficiently small correlation coefficient. The sampled time series can be mapped to a uniform distribution the interval $[0,1)$ by using equation $\ref{eq:1}$ with equation $\ref{eq:den1}$ where $\mu=0$:

\begin{equation}
\label{eq:map}
\zeta(v)=\frac{\mathrm{erf}(\frac{|v|}{\sqrt{2}\sigma})}{\mathrm{erf}(\frac{K}{\sqrt{2}\sigma})}
\end{equation}
$\zeta(v)$, defined in equation $\ref{eq:map}$, produces a set of random numbers between 0 and 1 when applied to a finite set of $v(t)$'s chosen at equally spaced times and solving equation $\ref{eq:1}$.

\subsection{Sampling Interval}\label{ssec:sampling}

Assuming the correlation function expressed in equation $\ref{eq:corr}$ holds for large enough time series, picking a sampling interval of $\bigtriangleup t = 10$ allows sampling for a series up to $t=10^{10}$, or equivalently, $10^9$ generated random values. However, for a larger time series and number of generated values $N$, the following requirement can be derived: \cite{KNUTH}

\begin{equation}
\label{eq:SI}
\frac{1}{2} \ln (N) < \bigtriangleup t
\end{equation}

\subsection{History Function Restriction}\label{ssec:history}

Using the map appearing in equation $\ref{eq:map}$, negative and positive history functions $\phi$ will generate the same numbers. Although the mapped time series is itself random, it would be useful from a RNG perspective to know that two different $\phi$'s produce a different set of random numbers. Thus, $\phi$ can be picked as either always positive or always negative to avoid two same sets of generated numbers for two different $\phi$'s. Furthermore, the sine function symmetry and shift properties also restrict $\phi$, since $|\sin(v)|=|\sin(n\pi \pm v)|$ for any integer $n$.  Thus, sets of numbers generated from different time series with different $\phi$'s should satisfy the following restriction:
\begin{equation}\label{eq:phiint}
\begin{split}
\phi_i=(0,1) \setminus \bigg( \phi_j= (\phi_i \pm \frac{n}{2\beta}) \cup (-\phi_i \pm \frac{n}{2\beta}) \bigg), \\ n \in [1,2,3,...), j \ne i
\end{split}
\end{equation}

\subsection{Problems with generation}

Generating random numbers with the scheme presented in this section is problematic for two reasons.
\begin{enumerate}
\item Generating numbers this way is slow. Sampling at every $\bigtriangleup t=10$ requires on the order $10^4$ computations for a single randomly generated number.

\item The map $\ref{eq:map}$ is hard to apply for random number generation because $v(t)$ may take the value of the maximum $K$. More precisely, when a truncated Gaussian is mapped with the $\zeta(v)$ map appearing in equation $\ref{eq:map}$, if a value where $v(t)=K$ happens to be sampled, it is mapped to exactly 1, which is not in the desired interval [0,1). This value can be individually removed from the set of generated random numbers but such a procedure may be inconvenient.
\end{enumerate}

\section{LSF-DDE-RNG}\label{sec:LSF}
\subsection{LSF scheme}\label{ssec:LSFs}

Although equation $\ref{eq:1}$ used as explained in Section \ref{sec:DDE-RNG} has no theoretical limit on the number of possible generated values, it is slow. In the last 10 years, different schemes have been presented to generate random numbers that use feedback laser mechanisms.\cite{LASER1,LASER2,LASER3} In such work, the measurements made from a laser system yield Gaussian distributed values which are used to generate random numbers. Two steps are employed: digit discarding and post-processing. The digit discarding involves only considering a certain amount of least significant bits (or digits) while completely discarding the others. The reasoning of Reidler et al. states that this allows fast random number generation, provided that the sampling rate is much slower than the chaos affecting the given least significant bits.\cite{LASER1} Also, Oliver et al.\cite{LASER3} state that ``the autocorrelation function of the captured time-series data is also affected by bit truncation, in such a way that residual correlations in the original dynamics are destroyed, and thus allowing for an increase in the rate of random bit generation."

For the post-processing, different schemes have been employed. Reidler et al. suggest taking differences between measured values to generate each random number as well as using an $xor$ operation on the least significant bits.\cite{LASER1} Oliver et al. suggest using an appropriate sampling rate after digit discarding. \cite{LASER3}

 Since the solution to equation $\ref{eq:1}$ is analogous to the measurement from these systems as the time series yields Gaussian distributed values, a similar scheme can be applied here, which will henceforth be referred to as LSF-DDE-RNG (Least Significant Figures).

	Although bit truncation can flatten a Gaussian distribution into an approximately uniform distribution,\cite{LASER3} this approximation breaks down as the number of generated values goes to infinity. This can be shown by calculating the expected probability for each set of possible bits resulting from the Gaussian distribution after digits (or bits) are discarded. Thus, the two-step digit discarding scheme can be applied with equation $\ref{eq:1}$ but a mapping function, equivalent to equation $\ref{eq:map}$, which maps values to a uniform distribution, should be used before discarding digits. The postprocessing for the LSF-DDE-RNG, presented here, is the use of a sufficient sampling rate.

\subsection{Revised Mapping Interval}
\label{ssec:map2}
Discarding a certain number of decimal digits from samples of numbers taken from a uniform distribution in [0,1) also yields a uniform distribution in the interval [0,1). A similar map to $\ref{eq:map}$ can be used to map samples from the solution to equation $\ref{eq:1}$ to a uniform interval, which does not involve the maximum $K$:

\begin{equation}
\label{eq:map2}
\xi(v) = {erf(\frac{|v|}{\sqrt{2}\sigma})}
\end{equation}

This equation allows mapping the solution from equation $\ref{eq:1}$ to a uniform distribution in the interval [0,$erf(\frac{K}{\sqrt{2}\sigma})]$ where $erf(\frac{K}{\sqrt{2}\sigma}$) is close to one. If more than zero decimal digits are discarded from samples taken from this interval, the samples will then be uniformly distributed in the interval (0,1]. The value for $\sigma$ computed as from equation $\ref{eq:sigma}$ has been revised and better follows the following relationship:

\begin{equation}
\label{eq:beta}
\sigma=\frac{1}{\sqrt(12.62677\beta-11.00613)}
\end{equation}


Equation $\ref{eq:beta}$ has negligible error if $\beta$ is picked between 20 and 50 and if digit discarding is used. In other words, equations $\ref{eq:1} $ and $\ref{eq:map2}$ with $\ref{eq:beta}$ can be used together for random number generation employing digit discarding.

\subsection{Digits Discarded}\label{subsec:DD}f
After mapping to a uniform distribution in the interval $[0,1)$ and discarding digits, the values produced are strictly positive. The appropriate empirical autocorrelation function can be expressed as

\begin{equation}
\label{eq:correlation2}
\rho = \frac{1}{N-1} \sum_{n=1}^{N-1}(\vel_n \vel_{n+1} - 0.25)
\end{equation}

Using the theoretical distribution of $\rho$, \cite{TESTU01} it is possible to test whether values behave as they should if they were truly drawn from a random sequence. This can be done by calculating the p-values from the autocorrelation for samples generated after discarding $m$ decimal digits. This also indicates whether or not there is correlation between successive values.

The p-values are shown in table $\ref{tab:correlation2}$. p-values are rounded to $10^{-2}$. Values between $0.01$ and $0.99$ are considered here to indicate negligible correlation between successive values. Here the number of generated values tested was $N=10^8$, and the values were sampled at every $\bigtriangleup t=0.001$. The time series solution was obtained as described in Section \ref{sec:chaotic}. The tabulated p-values suggest that discarding between 4 to 12 digits destroys correlation between successive values in the time series. Furthermore, digits 14 and above should not be used in random number generation. Double float data type precision was used.

\begin{table}\caption{\label{tab:correlation2} Autocorrelation p value after discarding $m$ decimal digits for $N=10^7$ }
\begin{ruledtabular}
\begin{tabular}{ccdddd}
 \mbox{$m$ no. of discarded digits} & \mbox{p-value} \\
\hline
1 & 0 \\
2 & 0 \\
3 & 1.00 \\
4 & 0.22 \\
5 & 0.10 \\
6 & 0.54 \\
7 & 0.32 \\
8 & 0.71 \\
9 & 0.06 \\
10 & 0.86 \\
11 & 0.38 \\
12 & 0.87 \\
13 & 1.00 \\

\end{tabular}
\end{ruledtabular}
\end{table}

Formally, digit discarding for a sample point $v(t)$ can be written as the function $DD(v)$ where

\begin{equation}
DD(v)=\frac{10^m v-floor(10^m v)}{10^m}
\end{equation}

In the above equation, $DD$ is the digit discarding function, $v(t)$ is a sample point from the time series, $m$ is the number of discarded digits and $floor$ represents the integer floor function.

\subsection{Lyapunov Exponent}

\begin{figure}[htbp]
\centering
\includegraphics[width=9.4cm]{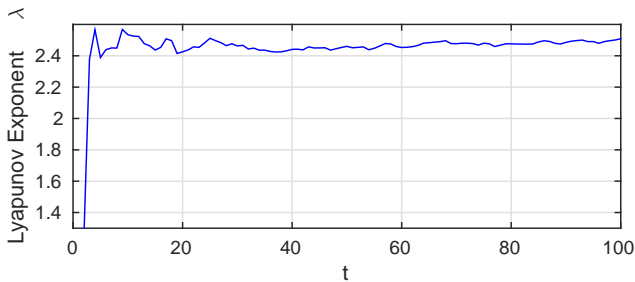}
\caption{\label{fig:Lyap} Maximal and averaged Lyapunov exponent for equation $\ref{eq:1}$ as computed by the methods of Breda and Van Fleck. \cite{BREDA} }
\label{fig:illustration}
\end{figure}

Chaotic dynamical systems are characterized by positive Lyapunov exponents (LE's).\cite{HAND} Using the methods provided by Breda and Van Fleck, \cite{BREDA} it was found that the maximal Lyapunov exponent $\lambda$ was 2.4496, as averaged over t=100 for equation $\ref{eq:1}$ and $\beta$=32.1357941. The computed $\lambda$ is shown in Figure \ref{fig:Lyap}. Reidler et al. stipulate that a requirement for random number generation should be that the ``sampling rate (clock period), is slow enough in comparison to the strength of the chaos, controlled by the spectrum of the Lyapunov exponents.''\cite{LASER1}  It remains an open question as to the direct relationship between the so-called clock period and the Lyapunov exponent, as implied by Reidler et al. However, we here speculate on a possible requirement between the amount of digits discarded $m$ and the maximal Lyapunov exponent $\lambda$:

\begin{equation}
\label{eq:liap}
e^{\lambda \bigtriangleup t}-1 > 10^{-m}
\end{equation}

Equation $\ref{eq:liap}$ holds in the case studied here, where equation $\ref{eq:1}$ is used with digit discarding, $-12 \leq m \leq -4$ and $\bigtriangleup t$=0.001. It is intuitive that if $\lambda$ were larger, less discarding of digits may be required (i.e. smaller $m$), and conversely, if $\lambda$ was negative, no random number generation could be achieved. The exponential functional form is intuitively suggested as it can quantify the divergence of initially separated trajectories. Anyhow, further work should   clarify whether equation $\ref{eq:liap}$ holds for other systems (e.g. different $\beta$, the experimentally realized laser systems, different DDE's). Also, besides equation $\ref{eq:liap}$, it remains unknown if there exists a quantitative statement of the claim from Reidler et al. quoted above. Although outside the scope of this paper, further study could determine to what extent the RNG laser feedback systems\cite{LASER1}\cite{LASER2}\cite{LASER3} are analogous to RNG generators using DDE's, such as the one presented in this work.

\subsection{Sampling Rate}

Although the autocorrelation function indicates no correlation between successive outputted values in the time series at every $\bigtriangleup t=0.001$ with $m= 4$ to $m=12$ discarded digits, values generated with this minimal sampling rate fail certain statistical tests of randomness. The values of the sampling rate had to be increased to  $\bigtriangleup t=0.002$ for the values to pass all required statistical tests. Figure $\ref{fig:scheme}$ shows all the steps needed to produce random numbers for the LSF-DDE-RNG. First, in Figure $\ref{fig:scheme}$ a), equation $\ref{eq:1}$ is solved with the Euler method as explained in Section $\ref{sec:DDE-RNG}$. In b), the mapping function $\ref{eq:map2}$ is used, as explained in Section $\ref{ssec:map2}$. It is used for every other (discrete) time series point, since $\bigtriangleup t=0.002$ was picked. In c), every two successive values from the mapped time series (red circles) yields a random number (black $\times$) as shown in d), after $m$=8 digits are discarded. In this case, two mapped values (red circles) must be used for one 10 digit random number ($2^{31}$ bits of resolution is standard), since keeping more than 9 digits from one number has been shown in table $\ref{tab:correlation2}$ to be undesirable (i.e. digits above the 4th and below the 14th are preferred). In other words, in this scheme, exactly four time series points yield one random number.

\begin{figure}[htbp]
\centering
\includegraphics[width=9.4cm]{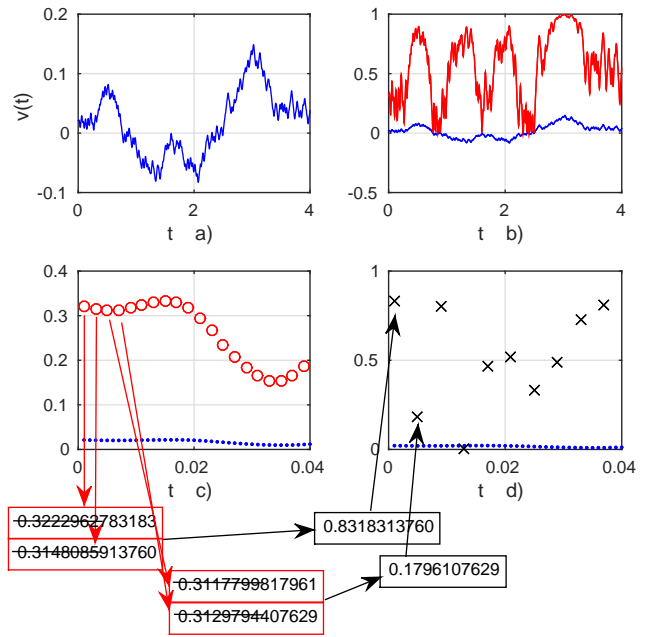}
\caption{\label{fig:scheme} a)-d) Time series solution for equation $\ref{eq:1}$ shown in blue, where $\beta=32.1357941$ and $\phi=0.8876641$. b) Equation $\ref{eq:map2}$ used on time series values (solid blue) to yield mapped values (solid red).  c) Same as b), but with a shorter time scale. The first four time series values (red circles) to be used in d) as random numbers (black $\times$), after m=8 digits are discarded, are explicitly shown in red boxes. The digit discarding is shown with a black strikethrough.}
\label{fig:scheme}
\end{figure}

\subsection{Statistical Tests of Randomness}\label{ssec:tests2}
Testing randomness of sets of numbers is quite involved. It requires checking both global randomness and local randomness. While testing these generated numbers, the null hypothesis is that all the generated numbers are truly random.  Many different tests have been proposed, for which, incidentally, the question of interdependency remains an open problem.\cite{SOTO} \citet{TESTU01} have compiled batteries of tests judged to be adequate in the testing of randomness. The most stringent of these batteries is TestU01's ``BigCrush". The tests involved, among others, the collision test, run test and the poker test. For example, the run test checks whether there are too few or too many monotonically increasing and decreasing subsequences. In the collision test, equally spaced intervals and the number of repeated values for a same bin, or collisions, are compared to the expected amount. In the poker test subsequences of values are treated as poker hands and are studied against expected hands.

Here, battery BigCrush is used from the TestU01 library, which tests random numbers with up to $2^{31}$ bits of resolution. This means a ten decimal random number is sufficient. Here the digits from 9 to 13 are used from sampled points in the time series, as shown in Figure $\ref{fig:scheme}$. In other words $m=8$ digits are discarded from every sampled point. A set of two values from the time series is needed for each 10 digit random number as using digits 9 to 13 yields five decimal digits. The sampling interval is picked to be $\bigtriangleup t=0.002$.

All tests from the BigCrush battery were passed when applied. $2.7\times10^{11}$ numbers generated from a single time series were verified for randomness using 160 statistical tests, including collision tests, run tests, and poker tests. The parameters used were $\beta=32.1357941$ and $\phi=0.8876641$, and the results of BigCrush have here been omitted due to their length. The criteria and specificity of the tests can be found in  \citet{TESTU01}. $10^8$ numbers generated from the LSF-DDE-RNG are illustrated in Figure $\ref{fig:illustration}$ a) by employing two consecutive numbers as $x$ and $y$ coordinates for $10^4$ points.

The same battery of stringent tests has not been applied for different $\beta$ and $\phi$, but from Lei and Mackey it is expected that other values of $(\beta, \phi)$, following the prescriptions of Sections \ref{ssec:map} and \ref{ssec:history}, could provide different, but also sufficiently random, sets. The highest precision for which $\beta$ and $\phi$ yield significantly different time series is unknown.

Finally we note that though some of the proposed feedback laser RNG's use the DIEHARD or NIST tests, \cite{LASER1,LASER2} the LSF-DDE-RNG with the above parameter $\beta$ and initial function $\phi$ is likely at least as random since the tests that were passed were much more stringent.\cite{TESTU01}

\begin{figure}[htbp]
\centering
\includegraphics[width=9.4cm]{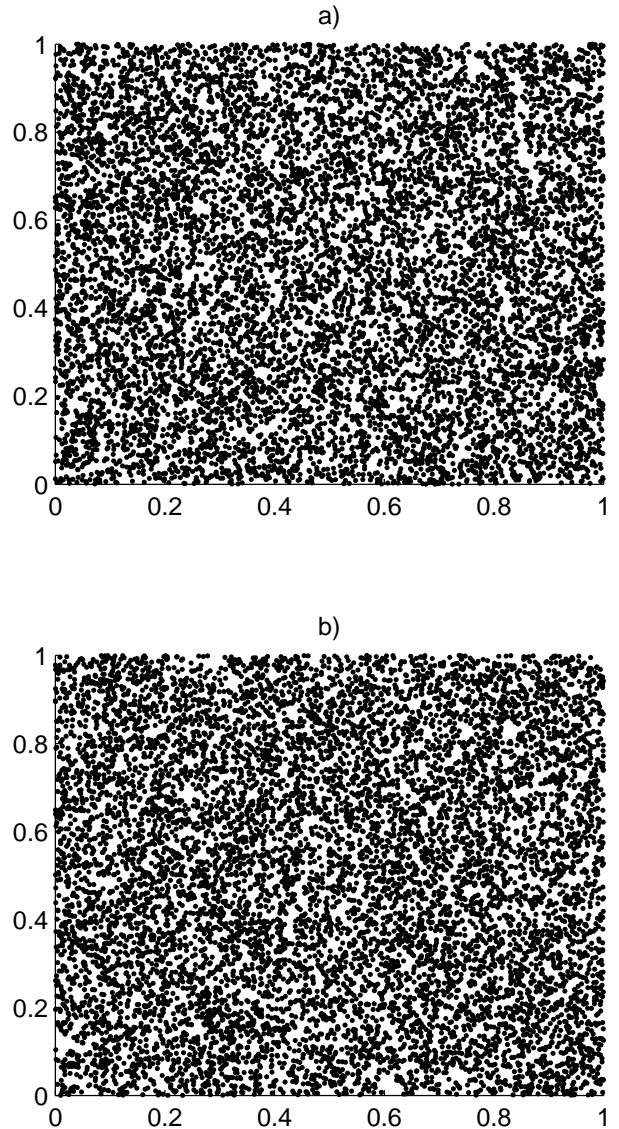}
\caption{ a) 10000x10000 points for which each two successive random numbers generated by the LSF-DDE-RNG are assigned to a set of $(x,y)$ coordinates. b) 10000x10000 points for which each two successive random numbers generated by MT19937 are assigned to a set of $(x,y)$ coordinates. }
\label{fig:illustration}
\end{figure}

\section{Benchmarking}\label{sec:benchmarking}

For benchmarking, a Multiple Recursive Generator was considered.\cite{ECUYER} A specific implementation already tested for good speed and randomness was used, MRG32k3a.  \cite{ECUYER}  Using C, $N=10^8$ numbers were generated with the LSF-RNG and MRG32k3a. MRG32k3a took about $4$ seconds for  $10^8$ values. The LSF-DDE-RNG generated $10^8$ random numbers in $26$ seconds, {  using a computer running Linux with a $2.50~$GHz Intel i5-2520M CPU}. The LSF-DDE-RNG's {  generation time} scales as N.

Although  the LSF-DDE-RNG generator is about 7 times slower than MRG32k3a, it has no practical period in its time series, while MRG32k3a has a period of $2^{191}$. However, it is advised to use many fewer than all the possible numbers generated from a given generator with a period, so the number of usable numbers are much less.\cite{ECUYER} More precisely, for the proposed LSF-DDE generator, although the initial function and subsequent states are finite (on the order of $10^{16000}$ unique states in the implementation presented in this paper), increasing the precision (e.g. long float instead of double float) would allow longer periods to prevent breakdown from computer accuracy should it ever be necessary to produce sequences larger than say $10^{10000}$ numbers. {Also, L'Ecuyer and Simard, in 2007, showed that many widely employed RNG's failed their ``Big Crush" battery of tests, and the generation time for $10^8$ random numbers for different generators was also reported, including  MRG32k3a.\cite{TESTU01} }

Figure $\ref{fig:illustration}$ b) shows $10^8$ numbers generated by the very widely used MT19937 ``Mersenne twister" (using Matlab software). Two successive numbers are used for a set of $x$ and $y$ coordinates of $10^4\times10^4$ points. One may look at the random numbers produced by LSF-DDE-RNG in Figure \ref{fig:illustration} a) and compare them to  Figure $\ref{fig:illustration}$ b). Although the Mersenne twister failed 2 tests from ``Big Crush", \cite{TESTU01} it is impossible to tell the quality of random number generators from visual inspection alone.

\section{Conclusion and Further Work}

It is intuitive that chaotic time series from a DDE could produce random numbers and the work detailed here proposes one such method. The digit discarding technique borrowed from feedback laser systems raises questions about the technique's relationship to the maximal Lyapunov exponent $\lambda$, for which a possible relation is speculated in equation $\ref{eq:liap}$.

Although it is the first of its kind, our proposed RNG, the LSF-DDE, produces random numbers on the same scale of quality, albeit slower, than its currently widely used counterparts MT19937 and MRG32k3a. It is not unimaginable that, like MRG's with LCG's, the ratio of speed to quality of our presented algorithm can be, in the future, dramatically increased due to improvements, perhaps in the underlying algorithm. It does nonetheless feature a fundamental difference from other popular software generators as it does not have a practical period.

Equation $\ref{eq:1}$ seems to be able to serve as a RNG, and similar equations may also be useful for RNG's. Equations $\ref{eq:dde-0}$ and $\ref{eq:dde-heav}$ have solutions with similar behaviour to solutions from equation $\ref{eq:1}$. And so, tests could be carried out to examine their usefulness as RNG's. Finally, the output quality of the proposed generator could be further checked by running tests for larger sequences of numbers. Different values of $\beta$ and $\phi$ could also be used to verify for similar randomness as these quantities are varied.

\begin{acknowledgments}
This work was supported by the Natural Sciences and Engineering Research Council (NSERC, Canada). We would like to thank Dimitri Breda for providing the scripts that were used to calculate the Lyapunov exponents and Tony Humphries, Erik Van Vleck, Joshua Lackman and Serhiy Yanchuk for their help.
\end{acknowledgments}


\bibliography{Random-2015-12-22}

\end{document}